\documentclass[aps, prd, superscriptaddress, nofootinbib, preprintnumbers, twocolumn, showpacs]{revtex4}
\usepackage{graphicx}
\usepackage{amsmath}
\usepackage{bm}
\def\fsl#1{\setbox0=\hbox{$#1$}                 
   \dimen0=\wd0                                 
   \setbox1=\hbox{/} \dimen1=\wd1               
   \ifdim\dimen0>\dimen1                        
      \rlap{\hbox to \dimen0{\hfil/\hfil}}      
      #1                                        
   \else                                        
      \rlap{\hbox to \dimen1{\hfil$#1$\hfil}}   
      /                                         
   \fi}                                         %
\newcommand{\tr}{\mbox{tr}}
\newcommand{\Tr}{\mathop{\mbox{Tr}}_{f,c,s}}

\newcommand{\sgn}{\mbox{sgn}}
\newcommand{\diag}{\mbox{diag}}

\newcommand{\VEV}[1]{\langle #1 \rangle}

\begin{document}
\title{Manifestation of Instabilities in Nambu-Jona-Lasinio type models}
\author{Michio Hashimoto}
\email{mhashimo@uwo.ca}
\affiliation{Department of Applied Mathematics, 
The University of Western Ontario, 
London, Ontario N6A 5B7, Canada}
\date{\today}
\preprint{UWO-TH-06/09}
\pacs{12.39.-x, 11.30.Qc, 26.60.+c}

\begin{abstract}
We study a Nambu-Jona-Lasinio (NJL) type model with two-flavor
quark matter in $\beta$-equilibrium.
It turns out that 
the system develops instabilities in the dispersion relations of 
the diquark fields, i.e.,
the velocity squared $v^2$ becomes negative 
in a certain region of the electron chemical potential.
The critical point is the same as that of the chromomagnetic instability.
The results imply the existence of spatially inhomogeneous 
diquark condensates in the genuine vacuum.
We also discuss gauge equivalent classes between 
the inhomogeneous diquark condensates and gluon condensates. 
\end{abstract}

\maketitle

\section{Introduction}

It was suggested a long time ago that quark matter might exist inside 
the central regions of compact stars~\cite{quark_star}.
At sufficiently high baryon density, cold quark matter is expected 
to be in a color superconducting state~\cite{CSC}.
For this reason, the color superconductivity has been
studied with a lot of interests.
(For reviews, see Refs.~\cite{review}.)

Bulk matter in the compact stars must be in equilibrium under
the weak interaction ($\beta$-equilibrium) and 
must be electrically and color neutral.
The electric and color neutrality conditions play a crucial role 
in the dynamics of the pairing between 
quarks~\cite{Iida:2000ha,Alford:2002kj,Steiner:2002gx,Huang:2002zd}.
For the compact stars the strange quark mass cannot be also neglected.
These give rise to a mismatch $\delta\mu$ between the Fermi momenta of 
the pairing quarks.

Recently, it has been revealed that the Meissner screening masses of 
gluons in the (gapped/gapless) two-flavor color 
superconducting phase (2SC/g2SC) turn into imaginary
in the supercritical regions of $\delta\mu$~\cite{Huang:2004bg,Huang:2004am}.
Such a chromomagnetic instability has been also found in 
the gapless color-flavor locked (gCFL) 
phase~\cite{Casalbuoni:2004tb,Alford:2005qw,Fukushima:2005cm}.

{}From the viewpoint of the quantum field theory,
it is quite natural to expect the existence of gluon condensates
in order to resolve the chromomagnetic instability~\cite{Gorbar:2005rx}.
Nevertheless, one might consider Nambu-Jona-Lasinio (NJL) type models 
without gluons as effective theories of QCD.
It is, of course, unlikely that all instabilities vanish away
in the NJL type models.
Then what is the alternative instability to the chromomagnetic one?

In this paper, we study two point functions of diquark fields,
which yield information about their low-energy properties.
We shall employ a linear representation for the diquark fields.
For the relation to an exponential parametrization 
(a decomposition to the radial and phase parts),
see e.g. Eq.~(31) in Ref.~\cite{Blaschke:2004cs}.
Physical quantities such as velocity are independent of 
the parameterizations, unlike off-shell ones.
Although we demonstrate only the two-flavor case,
our approach should be applicable also to the three-flavor quark matter.

The signal of the instability appears in the dispersion relations:
For the anti-red and anti-green diquark fields $\Phi^{r,g}$
(green-blue and blue-red quark pairing fields), which are 
the Nambu-Goldstone (NG) bosons
eaten by the 4-7th gluons if the theory is gauged, 
the velocity squared $v^2$ becomes negative 
in the region $\delta\mu > \Delta/\sqrt{2}$,
where $\Delta$ is a diquark gap\footnote{
As usual, we take the direction of the vacuum expectation value 
(VEV) of the color anti-triplet diquark fields to the anti-blue one.
}.
For the anti-blue diquark field $\Phi^b$ (red-green quark pairing field),
whose imaginary part is eaten by the 8th gluon after gauging
while the real part is left as it is,
gapless tachyons with $v^2 < 0$ emerge
in the region $\delta\mu > \Delta$, although the situation is involved
as we will mention below.
Each critical point is exactly the same as that for 
the chromomagnetic instability~\cite{Huang:2004bg,Huang:2004am}.

The instability $v^2 < 0$ has not yet been fully demonstrated 
in the earlier works~\cite{Reddy:2004my,Huang:2005pv,Hong:2005jv}. 
We also find that the behavior of $v^2$ for $\Phi^{r,g}$ 
in the whole region of $\delta\mu/\Delta$ is quite similar 
to that of the mass squared of the corresponding light plasmon 
obtained recently in Ref.~\cite{Gorbar:2006up}.
Furthermore, it turns out that the dispersion relation for $\Phi^b$
has two branches in the region $\delta\mu > \Delta$: 
One is normal while the other is tachyonic.
The tachyonic branch exists not only for the imaginary part of $\Phi^b$, 
but also for the real part.
It is rather surprising that the real part 
also has the instability $v^2 < 0$, because this field is not 
connected to the longitudinal mode of the 8th gluon and 
thus naively expected to be irrelevant to any instabilities.

The wrong sign of the velocity squared $v^2$ implies a wrong sign
for the spatial derivative terms of the diquark fields in 
the effective action, so that we expect the existence of 
{\it spatially inhomogeneous} diquark condensates
$\VEV{\Phi^\alpha}=\Delta^{\alpha}(\vec x)$, ($\alpha=r,g,b$)
in the genuine vacuum. 
Such non-uniform condensates have been studied in several contexts, e.g. 
``Larkin-Ovchinnikov-Fulde-Ferrell (LOFF) phase''~\cite{LOFF,Alford:2000ze},
``mixed phase''~\cite{Reddy:2004my},
``spontaneous NG current generation''~\cite{Huang:2005pv,Hong:2005jv},
``meson supercurrent state''~\cite{Kryjevski:2005qq,Schafer:2005ym}.
We also discuss gauge equivalent classes of 
the inhomogeneous diquark condensates. 

\section{Model}

We study the NJL model with two light quarks.
We neglect the current quark masses and
the $(\bar{\psi}\psi)^2$-interaction channel.
The Lagrangian density is then given by
\begin{equation}
  {\cal L} =
   \bar{\psi}(i\fsl{\partial}+\hat{\mu}\gamma^0)\psi
  +G_\Delta \bigg[\,(\bar{\psi}^C i\varepsilon\epsilon^\alpha\gamma_5 \psi)
           (\bar{\psi} i\varepsilon\epsilon^\alpha\gamma_5 \psi^C)\,\bigg],
 \label{Lag}
\end{equation}
where $\varepsilon$ and $\epsilon^{\alpha}$ are the totally 
antisymmetric tensors in the flavor and color spaces, respectively, 
i.e., $(\varepsilon)_{ij}=\varepsilon_{ij}$, $(i,j=u,d)$ and 
$(\epsilon^{\alpha})_{\beta\gamma}=\epsilon_{\alpha\beta\gamma}$, 
$(\alpha,\beta,\gamma=r,g,b)$ with $\varepsilon_{ud}=+1$ and 
$\epsilon_{rgb}=+1$.
The quark field $\psi$ is a flavor doublet and color triplet.
We also defined the charge-conjugate spinor 
$\psi^C \equiv C \bar{\psi}^T$ with $C = i\gamma^2\gamma^0$.
The whole theory contains free electrons,
although we ignore them in Eq.~(\ref{Lag}).
In $\beta$-equilibrium, the chemical potential matrix $\hat \mu$ 
for up and down quarks is 
\begin{equation}
  \hat{\mu} = \mu {\bf 1} - \mu_e Q_{\rm em}
              = \bar{\mu}{\bf 1} - \delta \mu \tau_3, 
\end{equation}
with ${\bf 1} \equiv {\bf 1}_c \otimes {\bf 1}_f$,
$Q_{\rm em} \equiv {\bf 1}_c \otimes \diag(2/3,-1/3)_f$,
and $\tau_3 \equiv {\bf 1}_c \otimes \diag(1,-1)_f$,
where
$\mu$ and $\mu_e$ are the quark and electron chemical potentials,
respectively.
(The baryon chemical potential $\mu_B$ is given by $\mu_B \equiv 3\mu$.)
We also defined $\bar{\mu} \equiv \mu - \delta\mu/3$ and 
$\delta \mu \equiv \mu_e/2$.
The subscripts $c$ and $f$ mean that the corresponding matrices act on 
the color and flavor spaces, respectively.
Hereafter, we abbreviate the unit matrices, ${\bf 1}$, 
${\bf 1}_c$ and ${\bf 1}_f$. 

In this paper, we neglect the color chemical potential $\mu_8$
because we employ the hard dense loop (HDL) approximation 
in the later analysis and 
the absolute value of $\mu_8$ in the 2SC/g2SC phase is suppressed 
by the quark chemical potential,
$\mu_8 \sim {\cal O}(\Delta^2/\mu)$~\cite{Gerhold:2003js}. 

Let us introduce the diquark fields
$\Phi^\alpha \sim i\bar{\psi}^C\varepsilon \epsilon^\alpha \gamma_5 \psi$.
We can then rewrite the Lagrangian density (\ref{Lag}) as
\begin{eqnarray}
  {\cal L} &=& \bar{\psi}(i\fsl{\partial}+\hat \mu \gamma^0)\psi
  - \frac{|\Phi^\alpha|^2}{4G_\Delta} \nonumber \\ &&
  - \frac{1}{2}\Phi^\alpha
     [i\bar{\psi}\varepsilon\epsilon^\alpha \gamma_5 \psi^C]
 - \frac{1}{2}
    [i\bar{\psi}^C\varepsilon\epsilon^\alpha\gamma_5 \psi]\Phi^{*\alpha}.
 \label{Lag_aux}
\end{eqnarray}
In the 2SC/g2SC phase,
we can choose the anti-blue direction without loss of generality, 
\begin{equation}
  \VEV{\Phi^r}=0, \quad  \VEV{\Phi^g}=0, \quad \VEV{\Phi^b}=\Delta ,
  \label{2SC}
\end{equation}
where the diquark condensate $\Delta$ is real.

We define the Nambu-Gor'kov spinor,
\begin{equation}
  \Psi \equiv \left(\begin{array}{@{}c@{}} \psi \\ \psi^C \end{array}\right) .
\end{equation}
The quark propagator $S$ in the Nambu-Gor'kov space is explicitly shown 
in Ref.~\cite{Huang:2004am}:
\begin{equation}
  S(K) = \left(
  \begin{array}{cc}
  G^+ & \Xi^- \\ \Xi^+ &  G^-
  \end{array}
  \right) ,
  \label{S}
\end{equation}
where
\begin{equation}
  G^\pm(K) \equiv \diag (G_\Delta^\pm,G_\Delta^\pm,G_b^\pm)_c,
\end{equation}
with
\begin{eqnarray}
  G_\Delta^\pm(K) &=& \phantom{+}
  \frac{(k_0\mp\delta\mu\tau_3)-E^\pm}
       {(k_0\mp\delta\mu\tau_3)^2-(E^\pm_\Delta)^2}\gamma^0 \Lambda_k^+
  \nonumber \\[2mm] &&
 +\frac{(k_0\mp\delta\mu\tau_3)+E^\mp}
       {(k_0\mp\delta\mu\tau_3)^2-(E^\mp_\Delta)^2}\gamma^0 \Lambda_k^-,\\[3mm]
  G_b^\pm(K) &=& \phantom{+}
  \frac{1}{(k_0\mp\delta\mu\tau_3)+E^\pm}\gamma^0 \Lambda_k^+
  \nonumber \\[2mm] &&
 +\frac{1}{(k_0\mp\delta\mu\tau_3)-E^\mp}\gamma^0 \Lambda_k^-,
\end{eqnarray}
and
\begin{equation}
  \Xi^\pm(K) \equiv \epsilon^b 
  \left(\begin{array}{cc}
    0 & \Xi_{12}^\pm \\ -\Xi_{21}^\pm & 0
  \end{array}\right)_f, 
\end{equation}
with
\begin{eqnarray}
  \Xi_{12}^\pm (K) &=&
  -i\Delta\left[\,\frac{1}{(k_0 \pm \delta\mu)^2-(E^\pm_\Delta)^2}
                   \gamma_5\Lambda_k^+ \right. \nonumber \\ && \qquad \left.
                 +\frac{1}{(k_0 \pm \delta\mu)^2-(E^\mp_\Delta)^2}
                   \gamma_5\Lambda_k^-\,\right], \\
  \Xi_{21}^\pm (K) &=&
  -i\Delta\left[\,\frac{1}{(k_0 \mp \delta\mu)^2-(E^\pm_\Delta)^2}
                   \gamma_5\Lambda_k^+ \right. \nonumber \\ && \qquad \left.
                 +\frac{1}{(k_0 \mp \delta\mu)^2-(E^\mp_\Delta)^2}
                   \gamma_5\Lambda_k^-\,\right]. 
\end{eqnarray}
Here $K^\mu \equiv (k_0,\vec k)$ denotes the energy-momentum four vector.
The matrices $G_{\Delta}^\pm$ and $G_b^\pm$ are $8 \times 8$ in 
the flavor-spinor space, 
while $\Xi_{12}^\pm$ and $\Xi_{21}^\pm$ having only spinor indices
are $4 \times 4$.
We also used the following notations,
$E^\pm \equiv |\vec k| \pm \bar{\mu}$, 
$E^\pm_\Delta \equiv \sqrt{(E^\pm)^2 + \Delta^2}$, and
$\Lambda_k^\pm \equiv \frac{1}{2}
 \left(1\pm\gamma^0\frac{\vec \gamma \cdot \vec k}{|\vec k|}\right)$.

\section{Two point functions for diquark fields}

Let us calculate the two point functions 
$\Gamma_{\alpha^{(*)}\beta^{(*)}}^{(2)}$
for the diquark channel $\Phi^{\alpha^{(*)}}$--$\Phi^{\beta^{(*)}}$.
In the fermion one-loop approximation, we find
\begin{subequations}
\label{2PF}
\begin{eqnarray}
\lefteqn{
  \Gamma^{(2)}_{\alpha\beta^*}(P) =  \Gamma^{(2)}_{\alpha^*\beta}(P) =
  -\frac{\delta^{\alpha\beta}}{4G_\Delta}
} \nonumber \\ &&
  -\frac{1}{2}\int\frac{d^4 K}{i(2\pi)^4}\Tr\Bigg[\,
   i\varepsilon\epsilon^\alpha\gamma_5 G^+(P+K)
   i\varepsilon\epsilon^\beta\gamma_5 G^-(K)\,\Bigg],
   \nonumber \\ && \label{2PF1} \\[-2mm]
\lefteqn{
  \Gamma^{(2)}_{\alpha\beta}(P) = 
} \nonumber \\ &&
  -\frac{1}{2}\int\frac{d^4 K}{i(2\pi)^4}\Tr\Bigg[\,
   i\varepsilon\epsilon^\alpha\gamma_5 \Xi^-(P+K)
   i\varepsilon\epsilon^\beta\gamma_5 \Xi^-(K)\,\Bigg],
   \nonumber \\ && \label{2PF2} \\[-2mm]
\lefteqn{
  \Gamma^{(2)}_{\alpha^*\beta^*}(P) = 
} \nonumber \\ &&
  -\frac{1}{2}\int\frac{d^4 K}{i(2\pi)^4}\Tr\Bigg[\,
   i\varepsilon\epsilon^\alpha\gamma_5 \Xi^+(P+K)
   i\varepsilon\epsilon^\beta\gamma_5 \Xi^+(K)\,\Bigg] ,
   \nonumber \\ && \label{2PF3}
\end{eqnarray}
\end{subequations}
where the ``Tr'' operation stands for the trace over 
the flavor ($f$), color ($c$), and spinor ($s$) spaces.
{}From the color-flavor structure, we easily find that the nonzero 
two point functions are
\begin{subequations}
\begin{align}
  &\Gamma^{(2)}_{r^* r}(P)=\Gamma^{(2)}_{r r^*}(P)=
  \Gamma^{(2)}_{g^* g}(P)=\Gamma^{(2)}_{g g^*}(P), \\
  &\Gamma^{(2)}_{b^* b}(P)=\Gamma^{(2)}_{b b^*}(P), \quad
  \Gamma^{(2)}_{b b}(P)=\Gamma^{(2)}_{b^* b^*}(P),
\end{align}
\end{subequations}
and the others are trivially vanishing.
It reflects the unbroken $SU(2)_c$ symmetry in the 2SC/g2SC phase.

For the anti-blue diquark field, it is convenient to introduce
two real scalar fields $\phi_{b1}$ and $\phi_{b2}$:
\begin{eqnarray}
  \Phi^{b}(x) &\equiv&
  \Delta+\frac{1}{\sqrt{2}}(\phi_{b1}(x)+i\phi_{b2}(x)), \\
  \Phi^{*b}(x) &\equiv&
  \Delta+\frac{1}{\sqrt{2}}(\phi_{b1}(x)-i\phi_{b2}(x)).
\end{eqnarray}
In the new basis, the two point function for the anti-blue diquark field
are diagonalized and then those for $\phi_{b1}$ and $\phi_{b2}$
are given by
\begin{eqnarray}
  \Gamma^{(2)}_{\phi_{b1}}(P) &=&
  \Gamma^{(2)}_{b^* b}(P) + \Gamma^{(2)}_{b b}(P),\\
  \Gamma^{(2)}_{\phi_{b2}}(P) &=&
  \Gamma^{(2)}_{b^* b}(P) - \Gamma^{(2)}_{b b}(P).
\end{eqnarray}

In the 2SC/g2SC phase, the color $SU(3)_c$ symmetry is spontaneously
broken down to $SU(2)_c$, so that one can expect to get
five NG bosons, i.e., two complex fields $\Phi^{r,g}$ 
(containing four real components) and
one real field $\phi_{b2}$.
When the $SU(3)_c$ symmetry is gauged,
the two diquark fields $\Phi^{r,g}$ are absorbed into 
the 4-7th gluons owing to the Anderson-Higgs mechanism, 
while $\phi_{b2}$ are eaten by the 8th gluon.
The real scalar field $\phi_{b1}$ is left as a physical mode.
However, there is a subtlety with respect to the number of 
the NG bosons\footnote{
The phenomenon with an abnormal number of 
NG boson was discovered in Ref.~\cite{abnormal_NG}
in the context of the dynamics of the kaon condensation.
}~\cite{Blaschke:2004cs,Ebert:2005fi,He:2005mp}.
If we introduce $\mu_8$ for the color neutrality
in the framework of the NJL type models, 
this $\mu_8$ explicitly breaks 
the color $SU(3)_c$ symmetry to $SU(2) \times U(1)$,
so that the diquark fields $\Phi^{r,g}$ acquire 
nonvanishing masses of the order of $\mu_8$~\cite{Ebert:2005fi,He:2005mp}.
We here emphasize that the Debye and Meissner screening masses 
have been calculated in the HDL approximation\footnote{
If we treat the suppressed terms as they are, it turns out that
three gluons of the unbroken $SU(2)_c$ subgroup acquire
peculiar Meissner masses proportional to $\Delta^2$~\cite{Rischke:2000qz}.
Otherwise, a subtraction scheme other than the vacuum subtraction one 
may be introduced~\cite{Alford:2005qw}.}~\cite{Rischke:2000qz,Huang:2004bg}.
For our purpose, it is then sufficient to employ the HDL approximation 
and thereby we can neglect $\mu_8$ because of 
$\mu_8 \sim {\cal O}\left(\frac{\Delta^2}{\mu}\right)$~\cite{Gerhold:2003js}.
Hence, within this approximation, 
the five NG bosons with linear dispersion relations appear 
in the spectrum, as we will see below.

We calculate the two point functions in the broken phase, so that 
we can eliminate the four-quark coupling constant $G_\Delta$
by using the gap equation, 
\begin{eqnarray}
  \frac{1}{4G_\Delta} &=& \phantom{-}
   2\int \frac{d^3 k}{(2\pi)^3}
   \left(\,\frac{1}{E_\Delta^+} + \frac{1}{E_\Delta^-}\,\right)
   \nonumber \\[2mm] &&
  -2\int \frac{d^3 k}{(2\pi)^3}
   \theta(-E_\Delta^- + \delta\mu)\frac{1}{E_\Delta^-}.
\end{eqnarray}
Within the HDL approximation, because of $|\vec k| \sim \mu$,
we can also justify the approximation,
\begin{equation}
  |\vec k + \vec p| \simeq k + p\xi,
  \qquad k \equiv |\vec k|, \qquad p \equiv |\vec p|,
  \label{E_pm_app}
\end{equation}
where $\xi$ is the cosine of the angle between the three momenta 
$\vec k$ and $\vec p$.
We then find approximately the trace over the Dirac matrices:
\begin{subequations}
\label{tr_D} 
\begin{align}
  \tr (\gamma_5 \gamma^0 \Lambda_{k+p}^\pm \gamma_5 \gamma^0 \Lambda_{k}^\pm)
  & \simeq 0, 
  &\tr (\Lambda_{k+p}^\pm \Lambda_{k}^\mp)
  & \simeq 0 , \label{tr_D1} \\
  \tr (\gamma_5 \gamma^0 \Lambda_{k+p}^\pm \gamma_5 \gamma^0 \Lambda_{k}^\mp)
  & \simeq  -2, 
  &\tr (\Lambda_{k+p}^\pm \Lambda_{k}^\pm)
  & \simeq  2. \label{tr_D2}
\end{align}
\end{subequations}
By using the prescription 
$k_0 \to k_0 + i\epsilon_+\sgn(k_0)$~(see, e.g., Ref.~\cite{Blaschke:2004cs}),
we can perform the integral over $k_0$ in Eqs.~(\ref{2PF1})--(\ref{2PF3}).
We do not report explicitly the results to save space,
but just mention that the relations, 
\begin{eqnarray}
& \Gamma_{r^*r}^{(2)}(P) = \Gamma_{g^*g}^{(2)}(P) \propto 
  P_\mu \Pi_{\pm}^{\mu\nu}(P) P_\nu,    \label{pi_4} \\[2mm]
& \Gamma_{\phi_{b2}}^{(2)}(P)  \propto 
  P_\mu \Pi_{88}^{\mu\nu}(P) P_\nu, 
  \label{pi_8}
\end{eqnarray} 
hold~\cite{Zarembo:2000pj,Rischke:2002rz},
where $\Pi_\pm^{\mu\nu}$ and $\Pi_{88}^{\mu\nu}$ are
the vacuum polarization tensors for the 4-7th and 8th gluons, 
respectively.

In the kinematic region $|p_0|, |p| \ll \Delta$,
we can explicitly perform the loop integral and 
find analytic expressions:
\begin{eqnarray}
\lefteqn{ \hspace*{-3mm}
  \Gamma^{(2)}_{r^* r}(P)=\Gamma^{(2)}_{g^* g}(P) } \nonumber \\
&& \hspace*{-5mm} 
 =\frac{\bar{\mu}^2}{\pi^2\Delta^2}\Bigg[\,
  \left(1+2\frac{\delta\mu^2}{\Delta^2}\right)\,p_0^2
  -\left(1-2\frac{\delta\mu^2}{\Delta^2}\right)\frac{p^2}{3}
\nonumber \\
&& \quad 
  -\theta(\delta\mu-\Delta)
    \frac{2\delta\mu\sqrt{\delta\mu^2-\Delta^2}}{\Delta^2}
    \left( p_0^2+\frac{p^2}{3}\right) \,\Bigg],
\label{2PF_rg}
\end{eqnarray}
\begin{eqnarray}
\lefteqn{
  \Gamma^{(2)}_{\phi_{b2}}(P) 
 =\frac{\bar{\mu}^2}{2\pi^2\Delta^2}\Bigg[\,
  p_0^2-\frac{p^2}{3}
} \nonumber \\
&&
 -\theta(\delta\mu-\Delta)\frac{\sqrt{\delta\mu^2-\Delta^2}}{\delta\mu}
  \left\{\,p_0^2
  -\frac{\delta\mu^2}{\delta\mu^2-\Delta^2}\frac{p^2}{3}
  \right. \nonumber \\ && \qquad
  \left.
  +\frac{\Delta^4\,p_0^2}{(\delta\mu^2-\Delta^2)^2}
  Q\left(\frac{\delta\mu}{\sqrt{\delta\mu^2-\Delta^2}}\frac{p_0}{p}\right)
   \,\right\}\,\Bigg] ,
\label{2PF_b2}
\end{eqnarray}
and
\begin{eqnarray}
\lefteqn{ \hspace*{-8mm}
  \Gamma^{(2)}_{\phi_{b1}}(P) =
 -\frac{2\bar{\mu}^2}{\pi^2}\Bigg[\,1-\theta(\delta\mu-\Delta)
  \frac{\sqrt{\delta\mu^2-\Delta^2}}{\delta\mu}
} \nonumber \\
&&
  \left\{\,1-\frac{\Delta^2}{\delta\mu^2-\Delta^2}
  Q\left(\frac{\delta\mu}{\sqrt{\delta\mu^2-\Delta^2}}\frac{p_0}{p}\right)
  \,\right\}\,\Bigg],
\label{2PF_b1}
\end{eqnarray}
where 
\begin{equation}
  Q(t) \equiv -\frac{1}{2}\int_0^1 d\xi
  \left(\,\frac{\xi}{\xi+t-i\epsilon_+}
         +\frac{\xi}{\xi-t-i\epsilon_+}\,\right), 
\end{equation}
i.e.,
\begin{equation}
Q(t) = \frac{t}{2} \ln \left|\frac{t+1}{t-1}\right| - 1
  - i \frac{\pi}{2}|t|\theta(1-t^2), \quad (\mbox{for real } t), 
\end{equation}
and
\begin{equation}
Q(t) = s \arctan \frac{1}{s}-1, \quad (\mbox{for imaginary } t \equiv is).
\end{equation}
For convenience, we described the results in the order of
$\Gamma_{\phi_{b2}}^{(2)}$ and $\Gamma_{\phi_{b1}}^{(2)}$.

\begin{figure}[tbp]
  \begin{center}
    \resizebox{0.45\textwidth}{!}{\includegraphics{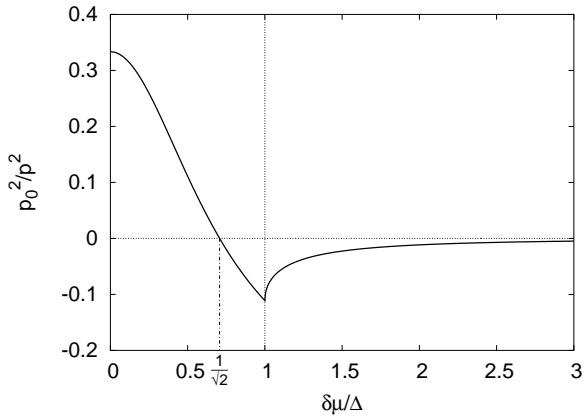}}
  \end{center}
\caption{The velocity squared for the anti-red and anti-green diquark fields
         (bold solid curve).
\label{fig1}}
\end{figure}

The zeros of $\Gamma^{(2)}_{r^* r}=\Gamma^{(2)}_{g^* g}$ 
in Eq.~(\ref{2PF_rg}) yield the dispersion relations for 
the anti-red and anti-green diquark fields,
\begin{equation}
  p_0^2 = \frac{p^2}{3}\frac{\Delta^2-2\delta\mu^2}{\Delta^2+2\delta\mu^2},
  \qquad (\delta\mu \leq \Delta),
  \label{disp_r1}
\end{equation}
and
\begin{equation}
  p_0^2 =
  -\frac{p^2}{3}
    \frac{\delta\mu-\sqrt{\delta\mu^2-\Delta^2}}
         {3\,\delta\mu+\sqrt{\delta\mu^2-\Delta^2}},
  \qquad (\delta\mu > \Delta) .
 \label{disp_r2}
\end{equation}
Therefore, the velocity squared $v^2 (\equiv p_0^2/p^2)$ is negative
for $\delta\mu > \Delta/\sqrt{2}$.
We also show the behavior of $v^2$ in Fig.~\ref{fig1}.
It is striking that Fig.~\ref{fig1} is quite similar to 
the behavior of the mass squared of the light plasmon 
for the 4-7th gluons obtained recently in Ref.~\cite{Gorbar:2006up}.

\begin{figure}[tbp]
  \begin{center}
    \resizebox{0.45\textwidth}{!}{\includegraphics{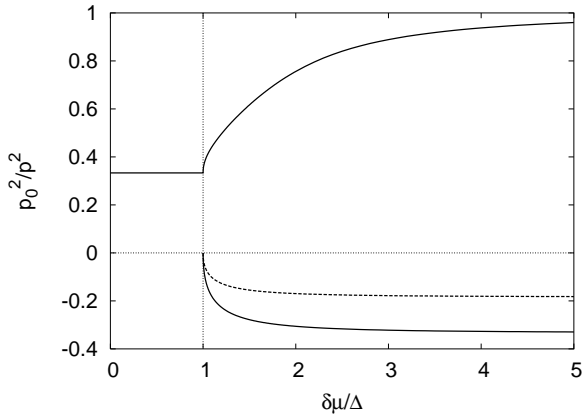}}
  \end{center}
\caption{The velocity squared for the anti-blue diquark field.
         The bold dashed and solid curves are for
         $\phi_{b1}$ and $\phi_{b2}$, respectively, where
         $\phi_{b1}$ corresponds to the quantum fluctuation of 
         the diquark condensate while $\phi_{b2}$
         does to one of the color $SU(3)_c$ phases.
         We ignored heavy excitations for $\phi_{b1}$.
\label{fig2}}
\end{figure}

The zeros of $\Gamma^{(2)}_{\phi_{b2}}$ in Eq.~(\ref{2PF_b2}) 
yield the dispersion relation for $\phi_{b2}$,
which corresponds to the longitudinal mode of the 8th gluon.
In the 2SC region, $\delta\mu < \Delta$, 
we obtain~\cite{Rischke:2002rz} 
\begin{equation}
  p_0^2 = \frac{p^2}{3}, \qquad (\delta\mu < \Delta),
 \label{disp_b2_2SC}
\end{equation}
which is exactly the same as the velocity squared
for the relativistic density sound wave.
On the other hand, in the g2SC region $\delta\mu > \Delta$, 
only in some limiting cases we can obtain analytic expressions.
In the vicinity of the critical point $\delta\mu=\Delta$, 
we find two solutions:
\begin{equation}
  \frac{p_0^2}{p^2} =
  \frac{1}{3}\left(\,1 + 
   \frac{4}{5}\frac{\sqrt{\delta\mu^2-\Delta^2}}{\delta\mu}\,\right), 
  \quad (\delta\mu=\Delta+0_+), 
\end{equation}
and
\begin{equation}   
  \frac{p_0^2}{p^2} =
 -\frac{3}{5}\frac{\sqrt{\delta\mu^2-\Delta^2}}{\delta\mu} ,
  \quad (\delta\mu=\Delta+0_+)\; .
\end{equation}
In the limit $\delta\mu \gg \Delta$, the two solutions approach
\begin{equation}
  \frac{p_0^2}{p^2}   \to 1, \quad \mbox{and} \quad
  \frac{p_0^2}{p^2} \to -\frac{1}{3} . 
\end{equation}
In a general case, we need to solve 
$\Gamma^{(2)}_{\phi_{b2}}(P)=0$ numerically.
We depict the results in Fig.~\ref{fig2} (bold curves).
There are two branches in $\delta\mu > \Delta$:
One is, say, an ultra-relativistic branch with $v^2 > 1/3$ and 
the other is a tachyonic one with $v^2 < 0$.
The behavior of the tachyonic branch is similar to that of 
the electric mode of the 8th gluon obtained recently in 
Ref.~\cite{Gorbar:2006up}.

The zeros of $\Gamma^{(2)}_{\phi_{b1}}$ in Eq.~(\ref{2PF_b1}) yield 
the dispersion relation for $\phi_{b1}$.
For $\delta\mu < \Delta$, however, there is no solution.
It implies that
the real part of the anti-blue diquark field acquires a heavy mass
of the order of $\mu$~\cite{Ebert:2004dr,Ebert:2005fi,He:2005mp}.
In such a case, a calculation beyond the HDL approximation is required
to obtain correctly the dispersion relation.
In this paper, we ignore this heavy excitation because
it is obviously irrelevant to the instability with our interests.
For $\delta\mu > \Delta$, we find an unexpected gapless tachyon.
In the vicinity of $\delta\mu = \Delta$, we find 
\begin{equation}
  \frac{p_0^2}{p^2} =
 -\frac{1}{3}\frac{\sqrt{\delta\mu^2-\Delta^2}}{\delta\mu},
  \qquad (\delta\mu=\Delta+0_+) .
\end{equation}
In the limit $\delta\mu \gg \Delta$,
the dispersion relation approaches
\begin{equation}
  \frac{p_0^2}{p^2} \simeq -0.184 .
\end{equation}
We depict the velocity squared for $\delta\mu > \Delta$ 
in Fig.~\ref{fig2} (dashed curve).
We do not deny a possibility that there may exist another branch with
a heavy excitation also in $\delta\mu > \Delta$.

We comment on the relation between the velocity squared $v^2$ for 
the NG bosons and the ratio of the Debye and Meissner screening masses 
($m_D$ and $m_M$).
One would expect $v^2 = m_M^2/m_D^2$~\cite{Son:1999cm},
because by the gauge principle 
the coefficients of $p_0^2$ and $\vec p{\;}^2$ 
for the kinetic terms of the diquark fields in 
the Ginzburg-Landau (GL) effective theory are inevitably proportional to 
$m_D^2$ and $m_M^2$, respectively.
(See, e.g., Refs.~\cite{Rischke:2000qz,Reddy:2004my}.)
In fact, this relation holds for the anti-red and anti-green diquark fields.
(Compare Eqs.~(\ref{disp_r1}) and (\ref{disp_r2}) with 
the ratio $m_M^2/m_D^2$ for the 4-7th gluons
shown in Refs.~\cite{Huang:2004bg,Huang:2004am}.)
However, it is not the case for $\phi_{b2}$, 
as one can check easily. 
Why does the naive expectation fail?

First of all, the 8th gluon couples to 
the blue quarks, while the anti-blue diquark field does not.
A difference then arises from this fact.
Note that the blue quarks affect the Debye screening mass $m_{D,8}$
of the 8th gluon. 
(Henceforth, for clarity, we use the notations $m_{D,8}$ and $m_{M,8}$
for the Debye and Meissner screening masses of the 8th gluon, respectively.)
If we employ the GL effective theory as in 
Refs.~\cite{Rischke:2000qz,Reddy:2004my} and choose appropriately 
the coefficients of the kinetic terms of the diquark fields
to reproduce correctly $m_{D,8}$ and $m_{M,8}$,
we obtain a wrong value of the velocity squared 
for the anti-blue diquark field. 
For example, in the 2SC region $\delta\mu < \Delta$ 
we find\footnote{
If we get rid of the blue quark contribution from $m_{D,8}$,
we can reproduce $v^2=1/3$ in $\delta\mu < \Delta$.}
$v^2 = 1/9$ inconsistently with Eq.~(\ref{disp_b2_2SC}).
In the framework of the GL effective theory, 
we cannot avoid this inconsistency.

The situation in the g2SC region $\delta\mu > \Delta$
is more involved.
The reason of the disagreement is not only for 
the blue quark contribution to $m_{D,8}$.
The point is that the 8th gluon and the anti-blue diquark field
now couple to two gapless modes 
and thereby the Feynman diagrams for the two point functions
$\Pi_{88}^{\mu\nu}$ and $\Gamma^{(2)}_{\phi_{b2}}$ 
contain the gapless quark loop.
Then a singularity appears at the point $(p_0,\vec p)=(0,\vec 0)$, 
i.e., the two point functions have the dependence of $p_0/p$
as in Eq.~(\ref{2PF_b2}).
This singularity arising from the gapless modes was recently 
revealed in Ref.~\cite{Gorbar:2006up}.
(The contributions of the massless blue quarks to 
the vacuum polarization tensor $\Pi_{88}^{\mu\nu}$ 
also have this singularity. In this sense, the disagreement 
is essentially caused by the existence of the singularity $p_0/p$
both in the regions $\delta\mu < \Delta$ and $\delta\mu > \Delta$.)
Note that Eq.~(\ref{pi_8}) can be written as
\begin{equation}
\Gamma^{(2)}_{\phi_{b2}}(P) \propto
  P_\mu \Pi_{88}^{\mu\nu}P_\nu = p_0^2 K_{88} - p^2 L_{88} -2 p_0 p M_{88} ,
 \label{Pi_para}
\end{equation}
where 
\begin{eqnarray}
\lefteqn{\hspace*{-5mm}
  \Pi_{88}^{\mu\nu}(P) \equiv
} \nonumber \\ &&
  (g^{\mu\nu}-u^\mu u^\nu
   + \frac{\vec p^{\,\mu}\vec p^{\,\nu}}{p^2}) H_{88}(P)
  + u^\mu u^\nu K_{88}(P) \nonumber \\[2mm] &&
  - \frac{\vec p^{\,\mu}\vec p^{\,\nu}}{p^2} L_{88}(P)
  +\left(u^\mu\frac{\vec p^{\,\nu}}{p}
        +u^\nu\frac{\vec p^{\,\mu}}{p}\right) M_{88}(P),
  \label{Pi8_HLKM}
\end{eqnarray}
with $g^{\mu\nu}=\diag(1,-1,-1,-1)$, $\vec p^{\,\mu} \equiv (0,\vec p)$,
and $u^\mu \equiv (1, 0, 0, 0)$. 
We can show that the blue quark contributions to 
$K_{88}$, $L_{88}$, and $M_{88}$ are cancelled out 
in the expression of Eq.~(\ref{Pi_para}).
The Debye and Meissner screening masses are
defined in the static limit $p_0=0$ and $p \to 0$;
\begin{eqnarray}
 m_{D,8}^2 &=& - K_{88}(p_0=0,p \to 0), \label {m_D} \\
 m_{M,8}^2 &=& -  H_{88}(p_0=0,p \to 0) . \label{m_M}
\end{eqnarray}
We also find
\begin{eqnarray}
  L_{88}(p_0=0,p \to 0) &=&   H_{88}(p_0=0,p \to 0), \label{L=H} \\
  M_{88}(p_0=0,p \to 0) &=& 0 . \label{M=0}
\end{eqnarray}
If no singularity, 
we could obtain $v^2=L_{88}/K_{88}=m_{M,8}^2/m_{D,8}^2$ from Eq.~(\ref{Pi_para})
and Eqs.~(\ref{m_D})--(\ref{M=0}).
(In fact, this is realized for $\Phi^{r,g}$.)
However, the function $\Gamma^{(2)}_{\phi_{b2}}$ depends on
$p_0/p$ in the g2SC region and 
the dispersion relation is defined at $p_0/p \ne 0$.
Therefore, the velocity squared $v^2$ disagrees 
with the ratio $m_{M,8}^2/m_{D,8}^2$ also in $\delta\mu > \Delta$,
even if we ignore the blue quark contribution to $m_{D,8}$.
In passing, the two point function $\Gamma^{(2)}_{\phi_{b1}}$ for 
$\phi_{b1}$ also includes the singularity $p_0/p$ in $\delta\mu > \Delta$ 
as shown in Eq.~(\ref{2PF_b1}), so that
a gapless tachyon emerges also in the $\phi_{b1}$ channel.

It should be interesting to construct a low-energy effective theory 
which can reproduce correctly $v^2$ for $\Phi^{r,g,b}$ if no gluons
and simultaneously the Debye and Meissner screening masses 
after gauging.
This issue is, however, beyond the scope of the paper.

\section{Gauge equivalent class of inhomogeneous diquark condensates}
\label{sec4}

In the previous section, we found the instability $v^2 < 0$ 
in the dispersion relations of the diquark fields.
It implies the existence of spatially inhomogeneous diquark condensates 
$\VEV{\Phi^{\alpha}} = \Delta^\alpha (\vec x)$, ($\alpha=r,g,b$)
in the genuine ground state.
On the other hand, when the gluons are introduced into the theory, 
the chromomagnetic instability $m_M^2 < 0$ is disclosed
and this phenomenon can be explained by 
the idea of gluon condensates as one of the self-consistent ansatz was 
demonstrated in Ref.~\cite{Gorbar:2005rx}.
Then what is the relation 
between the gluon condensates and the inhomogeneous diquark condensates?
We discuss below gauge equivalent classes between them.

We first consider the homogeneous gluon condensates 
$A_\mu^{a}$, ($a=1,2,\cdots,8$)
with the homogeneous diquark condensates $\Delta^\alpha$, ($\alpha=r,g,b$).
We do not specify here the direction of the homogeneous diquark 
condensates unlike Eq.~(\ref{2SC}).
We can show that only when there is no color magnetic field
$F_{jk}^a \equiv 0$, where $F_{\mu\nu}^{a}$'s are the field strengths,
\begin{equation}
  F_{\mu\nu}^{a} \equiv
  \partial_\mu A_{\nu}^{a} - \partial_\nu A_{\mu}^{a} +
  g f^{abc} A_{\mu}^{b} A_{\nu}^{c} =
  g f^{abc} A_{\mu}^{b} A_{\nu}^{c} ,
\end{equation}
with the QCD coupling constant $g$ and the structure constant $f^{abc}$,
the homogeneous diquark and gluon condensates,
\begin{equation}
  \Delta^\alpha, \quad A_0^a, \quad  A_j^a, 
  \label{gluonic}
\end{equation}
are gauge equivalent to the spatially inhomogeneous diquark condensates
and color chemical potentials,
\begin{equation}
\bm{\Delta}'(\vec x) = e^{i g A_j^{*} x^j} \bm{\Delta}, \quad 
\bm{\mu}'_{\rm col}(\vec x) = e^{-i gA_j x^j} \bm{\mu}_{\rm col} e^{i gA_j x^j},
\label{inhomo_Phi}
\end{equation}
in the NJL type models without the gluon fields
up to the tree level gluon potential\footnote{
The tree level potential for gluons is suppressed 
for reasonable values of the QCD coupling constant $g$, 
compared with the hard dense part of 
the one-loop corrections~\cite{Gorbar:2005rx}.
We can then ignore the tree term.}.
We here defined 
${\mathbf \Delta} \equiv (\Delta^r,\Delta^g,\Delta^b)^T$
and $\bm{\mu}_{\rm col} \equiv \mu_a T^a = g A_0^a T^a$
with the generators $T^a$ of $SU(3)$.
The gauge-singlet chemical potentials $\mu$ and $\mu_e$ are unchanged. 
When $F_{jk}^a \ne 0$, the phase of Eq.~(\ref{gluonic}) cannot be
described in the framework of the NJL type models.

The point is as follows:
In the context of the NJL type models, 
we can interpret the VEVs of 
the time components of gluons as the color chemical potentials, i.e., 
$\mu_a = g A_0^{a}$, by imposing the color neutrality 
conditions~\cite{Iida:2000ha,Buballa:2005bv}.
Therefore, the problem is only the existence of the space-component
VEVs $A_j^a$.
By using the spatially local $SU(3)$ matrix $U(\vec x)$,
\begin{equation}
  U(\vec x) \equiv e^{-i g A_j x^j} , \quad A_\mu \equiv A_\mu^{a}T^a,
\end{equation}
we can actually gauge away $A_j^a$ only when $F_{jk}^a \equiv 0$, because
\begin{eqnarray}
  A_j \to 
  A'_j (\vec x) &\equiv& \frac{i}{g} U(\vec x) \partial_j U^\dagger(\vec x)
  + U(\vec x) A_j U^\dagger(\vec x) , \label{A_j'} \\
&=&
  \frac{1}{2} F_{kj} x^k - \frac{ig}{3}[A_k,F_{\ell j}]x^k x^\ell +\cdots, 
\end{eqnarray}
where $F_{\mu\nu} \equiv F_{\mu\nu}^{a}T^a$.
In this case, the inhomogeneity of the color chemical potentials is 
connected with the existence of the color electric field.
If no color electric field $F_{0j}^{a} \equiv 0$,
the color chemical potentials are identical and thus homogeneous, 
because
\begin{eqnarray}
  A_0 \to 
  A'_0 (\vec x) &\equiv& U(\vec x) A_0 U^\dagger(\vec x) , \label{A_0'} \\
&=& A_0 + F_{k0}x^k -\frac{ig}{2}[A_j,F_{k0}]x^j x^k +\cdots .
\end{eqnarray}
However, when $F_{0j}^{a} \ne 0$
as in the gluonic phase demonstrated in Ref.~\cite{Gorbar:2005rx},
we cannot gauge away it.
(In the new basis, there still exists the color electric field,
$F_{0j} \to F'_{0j}(\vec x)=-\partial_j A_0'(\vec x) \ne 0$.)

We next consider the LOFF phase.

In two-flavor quark matter,
the 2SC-LOFF phase~\cite{Alford:2000ze}, 
\begin{equation}
\VEV{\Phi^r}=\VEV{\Phi^g}=0, \quad 
\VEV{\Phi^b}=\sum_{n=1}^N \Delta_n e^{2i\vec q_n \cdot \vec x},
\label{2SC-LOFF}
\end{equation}
has been studied.
The simplest case is $N=1$ and 
this single plane-wave LOFF phase 
can be described by a constant vector with the homogeneous 
diquark condensate~\cite{Giannakis:2004pf,Gorbar:2005rx,Gorbar:2005tx}. 
(See also Ref.~\cite{Fukushima:2006su}.)
The multiple plane-wave LOFF state with $N > 1$~\cite{Bowers:2002xr}
is generally gauge equivalent to 
the inhomogeneous radial part of the anti-blue diquark condensate
and the gradient of the phase part,
which corresponds to the inhomogeneous ``pure gauge'' term for 
the space component VEV of the 8th gluon.
The existence of the inhomogeneous condensates of the 8th gluon 
$\vec A^8(\vec x) \ne 0$ is suggested in Ref.~\cite{Gorbar:2006up}.
In this sense, the instability of the real part of $\Phi^b$ may be 
connected with the existence of $\vec A^8(\vec x) \ne 0$.

We may extend non-uniformity of the diquark condensates to 
all the directions,
$\VEV{\Phi^\alpha} = \Delta^\alpha (\vec x)$, $(\alpha=r,g,b)$,
as the negative velocity squared for $\Phi^\alpha$ implies.
We assume that the color chemical potentials $\mu_a$ are homogeneous. 
Notice that 
\begin{equation}
 \left( \begin{array}{@{}c@{}}
  \Delta^r(\vec x) \\ \Delta^g(\vec x) \\ \Delta^b(\vec x)
 \end{array}\right) = 
  \tilde{U}^*(\vec x)
 \left( \begin{array}{@{}c@{}}
  0 \\ 0 \\ \Delta(\vec x) e^{i\theta_0(\vec x)} 
 \end{array}\right),
\end{equation}
with 
\begin{equation}
  \tilde{U}^* (\vec x) \equiv 
  e^{i\theta_3(\vec x)\lambda^3} e^{i\theta_8(\vec x)\lambda^8}
  e^{-i\theta_2(\vec x)\lambda^2} e^{i\theta_5(\vec x)\lambda^5},
\end{equation}
where $\lambda^a$'s are the Gell-Mann matrices and we redefined
the diquark condensates by the radial part
$\Delta (\vec x) \equiv \sqrt{\sum_{\alpha=r,g,b}|\Delta^\alpha(\vec x)|^2}$,
and the phase parts $\theta_{0,2,3,5,8}(\vec x)$, i.e.,
$\tan \theta_2(\vec x) \equiv |\Delta^g(\vec x)|/|\Delta^r(\vec x)|$,
$\cos \theta_5(\vec x) \equiv |\Delta^b(\vec x)|/\Delta(\vec x)$, etc..
(The 2SC-LOFF state corresponds to 
$\theta_0(\vec x) \ne 0$, $\theta_{2,3,5,8}=0$ or
$\theta_8(\vec x) \ne 0$, $\theta_{0,2,3,5}=0$.)
By using the gauge transformation by $\tilde{U}^\dagger (\vec x)$,
we find that the inhomogeneous diquark condensates
and the homogeneous color chemical potentials,
\begin{equation}
 \Delta^\alpha (\vec x), \quad \mu_a, 
\end{equation}
in the NJL type models is equivalent to 
the inhomogeneous anti-blue diquark condensate with 
the inhomogeneous color chemical potentials,
\begin{equation}
  \Delta^{''b} (\vec x) = \Delta(\vec x)e^{i\theta_0(\vec x)}, \quad 
  \bm{\mu}''_{\rm col}(\vec x) =
  \tilde{U}^\dagger (\vec x)\bm{\mu}_{\rm col} \tilde{U}(\vec x),
\end{equation}
and with the inhomogeneous ``pure gauge'' terms,
\begin{equation}
  g A''_j(\vec x) \equiv 
  i\tilde{U}^\dagger (\vec x)\partial_j \tilde{U}(\vec x).
\end{equation}
One can check easily that all the color field strengths are vanishing
in the new basis.
In this case, the inhomogeneity of the color chemical potentials 
does not mean the existence of the color electric field
unlike the gluonic phase.

\section{Summary and discussions}

We studied the two point functions of the diquark fields
in the framework of the NJL type models. 
We found that the velocity squared $v^2$ 
for the anti-red and anti-green diquark fields
becomes negative in the region $\delta\mu > \Delta/\sqrt{2}$,
while for the anti-blue diquark field the gapless tachyons with $v^2 < 0$ 
appear in the region $\delta\mu > \Delta$.
The critical points of the instabilities $v^2 < 0$ are 
the same as those for the corresponding chromomagnetic instability.

It is remarkable that the dispersion relation of 
the anti-blue diquark field has two branches in the g2SC region 
$\delta\mu > \Delta$: 
One is normal while the other is tachyonic. (See Fig.~\ref{fig2}.)
Interesting is that the real part of the anti-blue diquark field
as well as the imaginary part
carries the instability $v^2 < 0$ despite the expectation
that this field should be heavy and have nothing to do with 
the chromomagnetic instability.

Our results imply the existence of the spatially inhomogeneous 
diquark condensates $\VEV{\Phi^{\alpha}} = \Delta^\alpha (\vec x)$, 
($\alpha=r,g,b$) in the genuine vacuum.
Although the 2SC-LOFF phase is taken to the anti-blue direction 
as in Eq.~(\ref{2SC-LOFF}),
we revealed that the spatially inhomogeneous condensates 
of the anti-red and anti-green diquark fields,
$\VEV{\Phi^r}=\Delta^r(\vec x)$ and $\VEV{\Phi^g}=\Delta^g(\vec x)$,
should play a very important role to resolve the instability 
at least in the 2SC region $\Delta/\sqrt{2} < \delta\mu < \Delta$, 
because the anti-blue diquark field does not have the instability $v^2 < 0$
in this region.
It is also consistent with the conclusions in 
Refs.~\cite{Gorbar:2005rx,Gorbar:2005tx}.

When we introduce the gluon fields into the theory,
the instability $v^2 < 0$ for the would-be NG bosons turns into
the corresponding chromomagnetic instability, while the instability 
for the real part of the anti-blue diquark field should be left as it is.
The negative Meissner mass squared implies the existence of 
the gluon condensates.
We thus discussed the relation between the inhomogeneous diquark
condensates and the gluon condensates in several cases.
As for the instability for the real part of the anti-blue diquark field 
in the region $\delta\mu > \Delta$, it may be connected with 
the existence of the inhomogeneous condensate of the 8th gluon
discussed in Ref.~\cite{Gorbar:2006up}.

There are some advantages in the language of the gluon condensates:
It is manifest whether or not the condensates are essentially homogeneous. 
The inhomogeneous diquark condensates can provide only 
the ``pure gauge'' terms even in the most general case, while
the gluon condensates are not restricted to the pure gauge.

Our approach casts a new light onto the problem of the chromomagnetic
instability.
It is also promising for the three-flavor case.

\begin{acknowledgments}
The author is very grateful to V.~A.~Miransky for fruitful discussions
and also would like to thank the APCTP for its hospitality.
This work was supported by the Natural Sciences and Engineering Research
Council of Canada.
\end{acknowledgments}


\begin{thebibliography}{99}

\bibitem{quark_star}
  D.~Ivanenko and D.~F.~Kurdgelaidze, Astrofiz. {\bf 1}, 479 (1965);
  Lett. Nuovo Cim. {\bf 2}, 13 (1969);
  N.~Itoh,
  Prog. Theor. Phys. {\bf 44}, 291 (1970);
  F.~Iachello, W.~D.~Langer and A.~Lande,
  Nucl. Phys. A {\bf 219}, 612 (1974).

\bibitem{CSC}
  D.~Bailin and A.~Love,
  Phys. Rept. {\bf 107}, 325 (1984);
  M.~Iwasaki and T.~Iwado,
  Phys. Lett. B {\bf 350}, 163 (1995);
  M.~G.~Alford, K.~Rajagopal and F.~Wilczek,
  Phys. Lett. B {\bf 422}, 247 (1998);
  R.~Rapp, T.~Sch\"{a}fer, E.~V.~Shuryak and M.~Velkovsky,
  Phys. Rev. Lett.  {\bf 81}, 53 (1998).

\bibitem{review}
  K. Rajagopal and F. Wilczek, in {\it Handbook of QCD}, 
  M.Shifman, ed., (World Scientific, Singapore, 2001), 
  [hep-ph/0011333];
  M.~G.~Alford,
  Ann. Rev. Nucl. Part. Sci. {\bf 51}, 131 (2001);
  D.~K.~Hong,
  Acta Phys. Polon. B {\bf 32}, 1253 (2001);
  S.~Reddy,
  Acta Phys. Polon. B {\bf 33}, 4101 (2002);
  T.~Sch\"{a}fer, hep-ph/0304281;
  D.~H.~Rischke,
  Prog. Part. Nucl. Phys. {\bf 52}, 197 (2004);
  M. Buballa, Phys. Rept. {\bf 407}, 205 (2005);
  M.~Huang,
  Int. J. Mod. Phys. E {\bf 14}, 675 (2005);
  I.~A.~Shovkovy, Found. Phys. {\bf 35}, 1309 (2005).

\bibitem{Iida:2000ha}
  K.~Iida and G.~Baym,
  Phys. Rev. D {\bf 63}, 074018 (2001)
  [Erratum-ibid. D {\bf 66}, 059903 (2002)].

\bibitem{Alford:2002kj}
  M.~Alford and K.~Rajagopal,
  JHEP {\bf 0206}, 031 (2002).

\bibitem{Steiner:2002gx}
  A.~W.~Steiner, S.~Reddy and M.~Prakash,
  Phys. Rev. D {\bf 66}, 094007 (2002).

\bibitem{Huang:2002zd}
  M.~Huang, P.~f.~Zhuang and W.~q.~Chao,
  Phys. Rev. D {\bf 67}, 065015 (2003).

\bibitem{Huang:2004bg}
  M.~Huang and I.~A.~Shovkovy,
  Phys. Rev. D {\bf 70}, 051501(R) (2004).

\bibitem{Huang:2004am}
  M.~Huang and I.~A.~Shovkovy,
  Phys. Rev. D {\bf 70}, 094030 (2004).

\bibitem{Casalbuoni:2004tb}
  R.~Casalbuoni, R.~Gatto, M.~Mannarelli, G.~Nardulli and M.~Ruggieri,
  Phys. Lett. B {\bf 605}, 362 (2005)
  [Erratum-ibid. B {\bf 615}, 297 (2005)].

\bibitem{Alford:2005qw}
  M.~Alford and Q. Wang,
  J. Phys. G {\bf 31}, 719 (2005).

\bibitem{Fukushima:2005cm}
  K.~Fukushima,
  Phys. Rev. D {\bf 72}, 074002 (2005).

\bibitem{Gorbar:2005rx}
  E.~V.~Gorbar, M.~Hashimoto and V.~A.~Miransky,
  Phys. Lett. B {\bf 632}, 305 (2006).

\bibitem{Blaschke:2004cs}
  D.~Blaschke, D.~Ebert, K.~G.~Klimenko, M.~K.~Volkov, and V.~L.~Yudichev,
  Phys. Rev. D {\bf 70}, 014006 (2004).

\bibitem{Reddy:2004my}
  S.~Reddy and G.~Rupak,
  Phys. Rev. C {\bf 71}, 025201 (2005).

\bibitem{Huang:2005pv}
  M.~Huang,
  Phys. Rev. D {\bf 73}, 045007 (2006).

\bibitem{Hong:2005jv}
  D.~K.~Hong,
  hep-ph/0506097.

\bibitem{Gorbar:2006up}
  E.~V.~Gorbar, M.~Hashimoto, V.~A.~Miransky and I.~A.~Shovkovy,
  Phys. Rev. D {\bf 73}, 111502(R) (2006).

\bibitem{LOFF}
  A. I. Larkin and Yu. N. Ovchinnikov, 
  Zh. Eksp. Teor. Fiz. {\bf 47}, 1136 (1964)
  [Sov. Phys. JETP {\bf 20}, 762 (1965)];
  P. Fulde and R. A. Ferrell, Phys. Rev. {\bf 135} A550 (1964).

\bibitem{Alford:2000ze}
  M.~G.~Alford, J.~A.~Bowers and K.~Rajagopal,
  Phys. Rev. D {\bf 63}, 074016 (2001).
  For a review, see, e.g.,
  R.~Casalbuoni and G.~Nardulli,
  Rev. Mod. Phys. {\bf 76}, 263 (2004).

\bibitem{Kryjevski:2005qq}
  A.~Kryjevski,
  hep-ph/0508180.

\bibitem{Schafer:2005ym}
  T.~Sch\"{a}fer,
  Phys. Rev. Lett. {\bf 96}, 012305 (2006).

\bibitem{Gerhold:2003js}
  A.~Gerhold and A.~Rebhan,
  Phys. Rev. D {\bf 68}, 011502(R) (2003);
  D.~D.~Dietrich and D.~H.~Rischke,
  Prog. Part. Nucl. Phys. {\bf 53}, 305 (2004).

\bibitem{abnormal_NG}
  V.~A.~Miransky and I.~A.~Shovkovy,
  Phys. Rev. Lett.  {\bf 88}, 111601 (2002);
  T.~Sch\"{a}fer, D.~T.~Son, M.~A.~Stephanov, D.~Toublan and 
  J.~J.~M.~Verbaarschot,
  Phys. Lett. B {\bf 522}, 67 (2001).

\bibitem{Ebert:2005fi}
  D.~Ebert, K.~G.~Klimenko and V.~L.~Yudichev,
  Phys. Rev. D {\bf 72}, 056007 (2005).

\bibitem{He:2005mp}
  L.~y.~He, M.~Jin and P.~f.~Zhuang,
  hep-ph/0504148.

\bibitem{Rischke:2000qz}
  D.~H.~Rischke,
  Phys. Rev. D {\bf 62}, 034007 (2000).

\bibitem{Zarembo:2000pj}
  K.~Zarembo,
  Phys. Rev. D {\bf 62}, 054003 (2000).

\bibitem{Rischke:2002rz}
  D.~H.~Rischke and I.~A.~Shovkovy,
  Phys. Rev. D {\bf 66}, 054019 (2002).

\bibitem{Ebert:2004dr}
  D.~Ebert, K.~G.~Klimenko and V.~L.~Yudichev,
  Phys. Rev. C {\bf 72}, 015201 (2005).

\bibitem{Son:1999cm}
  D.~T.~Son and M.~A.~Stephanov,
  Phys. Rev. D {\bf 61}, 074012 (2000)
  [Erratum-ibid. D {\bf 62}, 059902 (2000)].

\bibitem{Buballa:2005bv}
  M.~Buballa and I.~A.~Shovkovy,
  Phys. Rev. D {\bf 72}, 097501 (2005).

\bibitem{Giannakis:2004pf}
  I.~Giannakis and H.~C.~Ren,
  Phys. Lett. B {\bf 611}, 137 (2005); Nucl. Phys. B {\bf 723}, 255 (2005);
  I.~Giannakis, D.~f.~Hou and H.~C.~Ren, Phys. Lett. B {\bf 631}, 16 (2005).

\bibitem{Gorbar:2005tx}
  E.~V.~Gorbar, M.~Hashimoto and V.~A.~Miransky,
  Phys. Rev. Lett. {\bf 96}, 022005 (2006). 

\bibitem{Fukushima:2006su}
  K.~Fukushima,
  Phys. Rev. D {\bf 73}, 094016 (2006).

\bibitem{Bowers:2002xr}
  J.~A.~Bowers and K.~Rajagopal,
  Phys. Rev. D {\bf 66}, 065002 (2002).



\end{thebibliography}
\end{document}